\documentclass[aps,prl,twocolumn,showpacs,amsmath,amsfonts]{revtex4}
\usepackage{graphicx}
\begin{document}

\title{Electric field driven insulator-to-metal phase transitions}
\date{\today}

\author{M. Nardone} \affiliation{Department of Environment and Sustainability, Bowling Green State University, Bowling Green, OH 43403, USA}
\author{V. G. Karpov} \affiliation{Department of Physics and Astronomy, University of Toledo, Toledo, OH 43606, USA}

\begin{abstract}
We show that strong enough electric fields can trigger nucleation of needle-shaped metallic embryos in insulators, even when the metal phase is energetically unfavorable without the field. This general phenomenon is due to the gigantic induced dipole moments acquired by the embryos which cause sufficient electrostatic energy gain. Nucleation kinetics are exponentially accelerated by the field-induced suppression of nucleation barriers. Our theory opens the venue of field driven material synthesis. In particular, we briefly discuss synthesis of metallic hydrogen at standard pressure.

\end{abstract}

\maketitle

The topic of  metal-insulator transitions has long been established \cite{mott1990,imada1998} and remains active. Application aspects aside, it concentrates mostly on the underlying microscopic mechanisms, such as those induced by disorder (the Anderson transition), intra-atomic interaction (Mott-Hubbard), band crossing, and some others.  The common feature of these transitions is that they are driven by changes in some material parameter: degree of disorder, doping concentration, etc.

Here, we introduce the concept of electric field driven insulator-to-metal phase transitions.  They start with needle-shaped metal embryos forming in an insulator when the system is immersed in a strong enough field. We argue that, even when the bulk metallic phase as such is energetically unfavorable, increasing the field will eventually cause the transition to occur. Furthermore, assuming a fixed electric field, the final state of the system will be a uniformly metallic needle-shaped body.

This concept holds regardless of the microscopic mechanism of the transition (densification, crystallization, electron solvation, or others \cite{mott1990,imada1998}); it applies equally to solids and liquids.  For example, it predicts conductive needle-shaped crystallites forming in an insulating glass under strong enough fields. As another example, strong electric fields will trigger nucleation of liquid Si (metallic) needle-shaped inclusions in a semiconducting Si host, even at temperatures well below melting. A more provocative example, briefly discussed in this Letter, is the field-induced synthesis of metallic hydrogen under standard pressure.

The fact that symmetry-breaking electric fields can dramatically affect nucleation processes was recently realized \cite{karpov} while studying nucleation of highly conductive filaments in chalcogenide glasses of phase change memory \cite{ovshinsky1968}. Other related phenomena can include bias-induced metal-insulator transitions in resistive random access memory  \cite{hirose1976}, and dielectric breakdown in thin-film devices \cite{alam2002}.  Another category is non-photochemical laser induced nucleation \cite{garetz1996}. 

The theory herein takes a significant additional step by predicting that sufficiently strong fields can trigger transitions to states that would not be stable without the field; they will remain metastable upon field removal. As a possible candidate we mention bias-induced switching from insulating to highly conductive states, such as in vanadium dioxide (VO$_2$) \cite{stefanovich2000} and chalcogenide glasses \cite{ovshinsky1968}, where the conductive phase disappears upon field removal. Other known phenomena could be pertinent, such as {\it e. g.} dielectric breakdown in thin oxides. However, mainstream understanding  of the latter refers to a build-up of defects produced by stress, eventually forming the onset of a percolation path across the oxide \cite{alam2002}.  These types of mechanisms are beyond the present framework.

We start with a brief introduction to the electric field effect in classical nucleation theory (CNT). The free energy of a new particle in the presence of the field is,
\begin{equation}\label{eq:free}
F=A\sigma - \Omega\mu+F_E.
\end{equation}
Here $A$ and $\Omega$ are the particle surface area and volume, $\sigma$ is the coefficient of surface tension, and $\mu$ is the chemical potential difference between the two phases, taken to be positive when the bulk new phase is energetically favorable.  Eq. (\ref{eq:free}) does not specify the type of transition.  The electrostatic term has the form \cite{landau1984},
\begin{equation}\label{eq:electro}
F_E=-\frac{\varepsilon E^2 \Omega}{8\pi n},
\end{equation}
where $\varepsilon$  is the electric permittivity of the host insulating phase and the effect of particle geometry is embodied in the depolarizing factor, $n$.  For a sphere, $n = 1/3$, $A = 4\pi R^2$, and $\Omega = 4\pi R^3/3$.  In zero field, the maximum of $F(R)$ from Eq. (\ref{eq:free}) provides the nucleation barrier $W_0 = 16\pi\sigma^3/(3\mu^2)$ and radius $R_0 = 2\sigma/|\mu|$ with typical magnitudes near 1 eV and 1 nm.  Maintaining the assumption of spherical geometry, the field reduces the nucleation barrier and radius according to \cite{isard1977},
\begin{equation}\label{eq:classical}
W_{sph}=\frac{W_0}{\left(1+E^2/E_0^2\right)^2}\quad\mathrm{and}\quad R_{sph}=\frac{R_0}{1+E^2/E_0^2},
\end{equation}
where $E_0 = 2[W_0/(\varepsilon R_0^3)]^{1/2}$ is typically in the range of several MV/cm.

In CNT, $\mu$ must be positive in Eq. (\ref{eq:free}), implying a metastable host phase. Another important assumption is that of spherical symmetry. We show next how phase transitions are possible even for the case of energetically unfavorable bulk new phase (negative $\mu$) when a strong electric field is applied and the constraint of spherical shape is relaxed.
%

\begin{figure}[htb]
\includegraphics[width=0.48\textwidth]{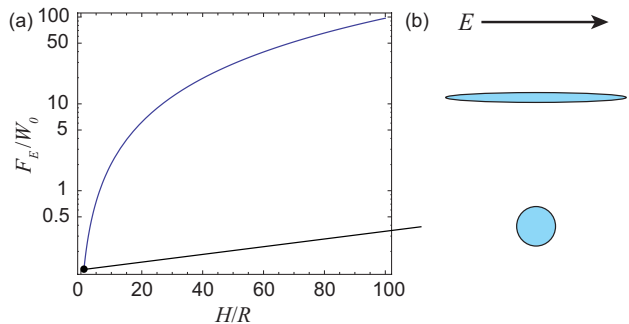}
\caption{(a) Absolute, normalized value of the electrostatic energy contribution, $F_E/W_0$, to the free energy as a function of nucleus aspect ratio; $H/R = 1$ corresponds to a sphere.  A value of $E/E_0 = 0.25$ and nucleus volume of $4\pi R_0^3/3$ were used for comparison purposes.  (b) Surface plots of the electric potential and streamlines of the electric field (shown in red) illustrate the greater electrostatic energy reducing effect of elongated metallic nuclei versus spheres.  An elongated nucleus concentrates the field at its tips, similar to the lightning rod effect, possibly triggering further nucleation events. \label{Fig:electric}}
\end{figure}

In our concept, the free energy of Eq. (\ref{eq:free}) has two degrees of freedom: spherical symmetry is broken and, when the field is sufficiently strong, needle-like conductive particles aligned with the field become energetically favorable.  That can be understood by comparing the electrostatic energy contribution of a sphere to that of a prolate spheroid of height $H$ and radius $R$, for which the depolarizing factor is \cite{landau1984},
\begin{equation}\label{eq:depol}
n = (R/H)^2[\ln{(2H/R)}-1]\equiv(R/H)^2L.
\end{equation}
Considering particles of equal volume, the electrostatic contribution is greater for a prolate spheroid by a huge factor of approximately $(H/R)^2 \gg 1$ (see Fig. \ref{Fig:electric}).  Physically, this enhancement is due to a large induced electric dipole in the needle-shaped particle.  Once created, it will act as a lightning rod, concentrating the field and triggering further nucleation.

\begin{figure}[htb]
\includegraphics[width=0.48\textwidth]{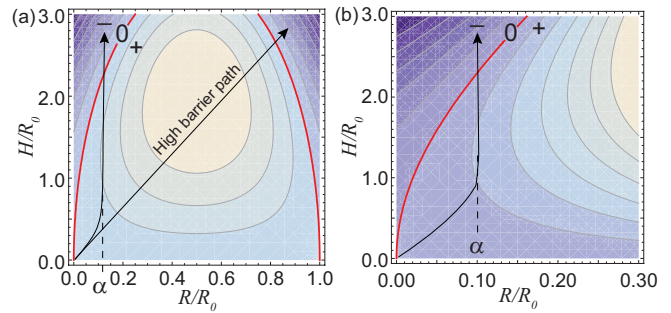}
\caption{Free energy landscape of FIMS as a function of nucleus radius, $R/R_0$, and height, $H/R_0$ [from Eq. (\ref{eq:freecyl}) with $E/E_0 = 0.25$].  Contour spacings are $F/W_0\sim 0.1$.  Regions of positive and negative free energy are separated by the zero contour (red).  In (a), the new phase is stable in the bulk [negative sign in Eq. \ref{eq:freecyl}], while in (b) the new phase would be unstable [positive sign in Eq. \ref{eq:freecyl}] if the electric field were not present (note the difference in scale).  The contours show that when the field is strong enough, nucleation pathways with much lower barriers become available for elongated embryos, regardless of bulk phase stability without the field. \label{Fig:free}}
\end{figure}

The exact shape of the elongated nucleus is not known, but modeling with either spheroidal or cylindrical particles leads to differences only in numerical coefficients \cite{karpov}.  We opt for the mathematically concise cylinder shape with $A = 2\pi RH$, $\Omega = \pi R^2H$, and free energy,
\begin{equation}\label{eq:freecyl}
F_{cyl}=\frac{W_0}{2}\left(\frac{3RH}{R_0^2}\pm \frac{3R^2H}{R_0^3}-\frac{E^2}{E_0^2}\frac{H^3}{R_0^3}\right),
\end{equation}
with the approximation $n \approx (R/H)^2$ for $H\gg R$ in Eq. (\ref{eq:depol}).  As illustrated in Fig. \ref{Fig:free}, the free energy landscape exhibits a range of low nucleation barriers at small $R$.

In the second term of Eq. (\ref{eq:freecyl}), we allow $\mu$ (included in $R_0$) to be negative for a new phase that is energetically unfavorable in the bulk.  Eq. (\ref{eq:freecyl}) predicts needle-shaped second phase particles to be energetically favorable provided that,
\begin{equation}\label{eq:aspect}
\frac{H}{R}>\sqrt{3\left(1+\frac{R_0}{R}\right)}\frac{E_0}{E}.
\end{equation}
Figs. \ref{Fig:free} shows, indeed, that in a strong field the system lowers its energy more easily by forming elongated particles, regardless of the sign of $\mu$.  That is the general mechanism by which the field can drive the transition.

While post-nucleation growth is beyond the present scope, the final state can be readily described in the same framework. If the final state is a stand-alone metallic body in the same field $E$, then the obvious modifications to Eqs. (\ref{eq:freecyl}) and (\ref{eq:aspect}) will be as follows: $\mu$ will have the meaning of the chemical potential of that metal, $\sigma$ will stand for its surface tension, and $\varepsilon$ should be set to unity assuming that the body is in vacuum. Therefore, the transformation will result in a uniformly metallic needle-shaped body.

Eq. (\ref{eq:freecyl}) suggests that nuclei with $R\rightarrow 0$ are most favorable. Realistically, $R$ must be greater than some minimum value determined by extraneous requirements, such as sufficient conductivity to support a large dipole energy or mechanical integrity.  Based on relevant data, it was estimated \cite{karpov} that $R_{min} = \alpha R_0$, where $\alpha\sim 0.1$ is a phenomenological parameter.  That puts $R_{min}$ in the range of molecular size.  The free energy in the region $R < R_{min}$ is substantially larger than described by Eq. (\ref{eq:freecyl}) because the energy reducing effect of the electric field cannot be manifested by such thin particles; this can be approximated by a potential wall.  With the latter in mind, the maximum of the free energy in Eq. (\ref{eq:freecyl}) (with $R = \alpha R_0$) yields the nucleation barrier \cite{karpov},
\begin{equation}\label{eq:barriercyl}
W_{cyl}=W_0\frac{\alpha^{3/2}E_0}{E}.
\end{equation}
The associated critical aspect ratio is, $H_c/R_{min} = E_0/(E\alpha^{1/2}) \gg 1$.

\begin{figure}[htb]
\includegraphics[width=0.4\textwidth]{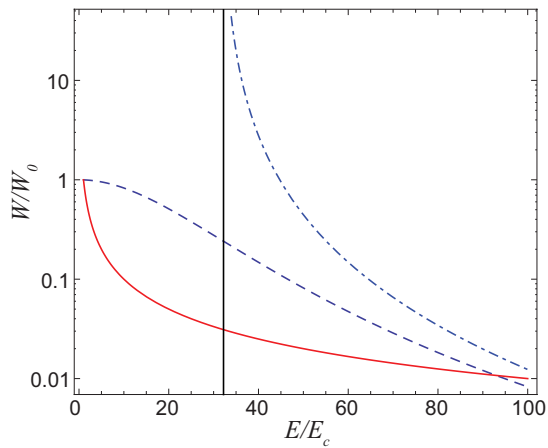}
\caption{Normalized nucleation barrier, $W/W_0$, as a function of the applied field, $E$, relative to the critical field $E_c$.  The barrier for nucleation of elongated particles (red, solid line) [Eq. (\ref{eq:barriercyl})] is compared to that of spheres [Eq. (\ref{eq:classical})] for the cases of: a stable new phase [$\mu > 0$ in Eq. (\ref{eq:free})] (blue, dashed line); and an unstable new phase [$\mu < 0$ in Eq. (\ref{eq:free})] (blue, dash-dot line).  Nucleation of elongated particles is highly favored across the entire field range ($E_c < E < Ec/\alpha^2$), where $\alpha = 0.1$.  Nucleation in the region $E > E_c/\alpha^2$ is uncertain due to the requirement of ultra-small nuclei.  In the region to the left of the vertical line, the field can drive the transition to a phase that would be unstable without the field; that is the region where otherwise unobtainable materials may be synthesized.\label{Fig:barriers}}
\end{figure}

The barrier of Eq. (\ref{eq:barriercyl}) is suppressed when $E > E_c = \alpha^{3/2}E_0$.  Correspondingly, nucleation of needle-shaped particles is vastly accelerated by electric fields under which spherical particle nucleation would be practically unaffected [cf. Eq. (\ref{eq:classical})], as illustrated in Fig. \ref{Fig:barriers}.  Moreover, for the case of $\mu < 0$, there exists a field range ($E < E_0$) where spherical nucleation is not possible and nucleation only occurs via needle-shaped embryos (left of the vertical line in Fig. \ref{Fig:barriers}).

%

It should be noted that while significant as the phase change driver, the electric polarization here remains small with respect to the total charge distribution in the needle-shaped embryo.  Indeed, the charge moved to the embryo ends is estimated as $q_E\sim EH_c^2$, with $H_c$ given below Eq. (\ref{eq:barriercyl}).  That should be compared to the total charge of the embryo $q\sim eH_cR_{min}^2/a^3$, where $e$ is the electron charge and $a$ is the characteristic interatomic distance:  $q_E/q\sim E_0/(E_{at}\sqrt{\alpha})\ll 1$, where $E_{at}\sim e/a^2\sim 10^9$ V/cm is the characteristic atomic field.  A more accurate analysis in Ref. \cite{landau1984} (p. 17) shows that the ratio $q_E/q$ is further reduced by a factor of  $1/L$, where $L \gg 1$ is defined in Eq. (\ref{eq:depol}).

Another comment concerns field-induced transitions in the vicinity of a bulk phase transition determined by the critical temperature ($T_c$), pressure, or concentration, such as {\it e. g.} the bulk phase transition between the insulating and conductive phases of VO$_2$  at $T_c=340$ K.  Using the standard approximation $\mu = \mu_0(1-T/T_c)$, where $\mu _0$ is the chemical potential difference between the two phases at zero temperature,  results in the corresponding renormalization [cf. Eq. (\ref{eq:classical})],
\begin{equation}\label{eq:renorm}
W_0\propto {(1-T/T_c)^{-2}},  R_0\propto (1-T/T_c)^{-1},E_0\propto (1-T/T_c)^{1/2}.\nonumber
\end{equation}
Because $R_{min}$ is determined by the microscopic structure and remains practically independent of $T$, we observe that $\alpha\propto (1-T/T_c)$. As a result, the barrier $W_{cyl}$ is temperature independent. That conclusion is in striking contrast to the prediction of CNT that the nucleation barrier is strongly temperature dependent, $W_0\propto \mu^{-2}\propto (1-T/T_c)^{-2}$.  Thus, we observe that field-induced nucleation becomes exponentially more effective than the classical nucleation of spherical particles in the proximity of bulk phase transition. It can dominate even under relatively weak fields $E>E_{00}\alpha _{00}^{3/2}(1-T/T_c)^2$, where $E_{00}$ and $\alpha_{00}$ are the zero-temperature values of $E_0$ and $\alpha$. This effect can be rather substantial.  For example, $1-T/T_c\approx 0.2$ for the case of VO$_2$ at room temperature yields $E>100$ V/cm.
We note that needle-shaped nuclei in polycrystalline VO$_2$ have been observed \cite{nucrad}, which can be attributed to nucleation in the internal fields induced by the grain boundaries.

Field driven phase transitions would also be enhanced in laser or dc fields that are sufficiently strong to ionize the material.  Indeed, that process would generate free charge carriers, thereby increasing the system polarizability and its related trend toward the transformation.


As a provocative example, consider next the synthesis of metallic hydrogen (MH).  Predicted by Wigner and Huntington \cite{wigner1935} in 1935, solid MH has not yet been observed under static pressures of up to 342 GPa \cite{loubeyre2002, narayana1998}.  Dynamic compression beyond 200 GPa has also been employed \cite{boriskov2010, hicks2009}.  The only direct evidence thus far was the brief observation \cite{weir1996} of a highly conductive liquid phase under a shockwave pressure of 140 GPa and temperature around 3000 K.  We will now attempt a rough estimate of the electric field range under which MH could be synthesized under standard pressure.

We use $\mu\sim 0.1$ Ry/atom \cite{brovman1972a} for the difference in chemical potential between the molecular and monatomic phases, and $\sigma\sim 1$ Ry/atom (as a rough order of magnitude estimate for significantly different structures), yielding $R_0\sim 10$ $\textrm{\AA}$ and $W_0\sim \mu R_0^3\sim 10^3$ eV.  Assuming $\alpha= 0.1$, and $\varepsilon\sim 1$ \cite{boriskov2010}, we obtain the critical field $E_c = \alpha^{3/2}E_0\sim 10^7$ V/cm, or equivalent laser intensity $I_c\sim 10^{12}$ W/cm$^2$.  Therefore, the practical window for field-induced synthesis of MH is $10^7 < E \ll 10^9$ V/cm.  That field range could be made even lower in the proximity of the bulk phase transition (e.g. close to the critical pressure).
Investigation could also be conducted with hydrogen rich alloys, such as CH$_4$ (or other paraffins) or SiH$_4$(H$_2$)$_2$ \cite{eremets2008}.  We note that while MH particles can be field-induced according to our estimates, they will remain metastable and will exist for only a finite time upon field removal due to the inequality in bulk chemical potentials ($\mu < 0$) and the metastability barrier of $\sim 1$ eV \cite{brovman1972b}.

We shall end by briefly mentioning the possibility of reverse, metal-to-insulator transformations when the electric potential (rather than the electric field) is kept constant. Here we limit our arguments to an analogy with the well-known elementary physics problem of a capacitor with plates partially immersed in water. When it is disconnected from the voltage source $U$, its charge $Q$ (hence, electric induction) is conserved and the energy $Q^2/2C$ tends to decrease via the increase in capacitance $C$; this is achieved by pulling in the more polarizable substance (water). This is analogous to creating more polarizable metal particles in the above analysis. However, the capacitor will push water away when it is connected to the voltage source in order to decrease its energy $CU^2/2$. In terms of our consideration, this corresponds to eliminating the more polarizable metal phase. 

In conclusion, we have shown how a symmetry-breaking electric field can drive an insulator-metal phase transition in any dielectric through nucleation of needle-shaped particles, eventually transforming it to a uniformly metallic elongated body.  We determined the conditions under which such phenomena are possible and their corresponding transformation rates. From a practical standpoint, the concept of field-induced phase transitions can stand as a unique pathway for synthesis of materials that may be otherwise unobtainable.

\end{document}